# Using stories to bridge the chasm between perspectives: How metaphors and genres are used to share meaning

**Emily Keen**
Department of Computing and Information Systems
University of Melbourne
Melbourne, Australia
Email: emily.keen@unimelb.edu.au

**Associate Professor Simon Milton**
Department of Computing and Information Systems
University of Melbourne
Melbourne, Australia
Email: Simon.Milton@unimelb.edu.au

**Dr Rachelle Bosua**
Department of Computing and Information Systems
University of Melbourne
Melbourne, Australia
Email: Rachelle.Bosua.unimelb.edu.au

## Abstract

Natural language is complex in structure, and contains considerable detail. All instances of language serve the purpose of making sense of experience and the intent of actors. Language conveys an actor's personal reference to goals, responsibility, and values. In this paper we consider how actors from distinct perspectives communicate when they have share common goals and intent. We have observed the planning of a community and cultural event where actors are variously responsible for management and for artistic merit.

Specifically, we consider actors' use of language as a tool to span perspectives, and how functional discourse analysis tools and techniques enable a deeper interpretive understanding of the layers of discourse when derived from a rich context. We will also illustrate patterns we have found in the use of discourse and actors' ability to bridge the reasoning and logical gap between distinct perspectives through references to metaphors and genres.

## 1  Introduction

Metaphor, as a phenomenon, involves both conceptual mappings and individual linguistic expressions (Lakoff, 1992). If figurative language is used in social setting and meetings, actors cannot interpret expressions and stories literally, but instead refer to existing knowledge and personal perspectives to form personally meaningful idiom and create personal 'real' meaning or interpretive literal meaning from such metaphors as "people that are going to put the *cart before the horse*, that are going to be good at running events but not know why they're running an event".

This paper looks at the value of metaphors and story genres to bridge perspectives and create common imagery, when there are distinct perspectives and reference domains between actors in a social setting.

In this paper we consider the value of metaphors for the purpose of reasoning, thoughts and expression. As indicated by (Lakoff, 1992) a metaphor is not a matter of language, but of thought and reason, we therefore consider actors' approaches to reasoning and communication when faced with multiple actors and evidence of non-literal language.

If we are to map the relationships between target and reference domain concepts, then the language is secondary, in that it sanctions the use of source domain language and inferences patterns for target domain concepts, formation and meaning by stating the generalizations and commonalities. In this paper we are in particular interested in examining the roles of metaphors to investigate the formation of ontology concepts and ontology concept mappings.

The following section will discuss the value of interpreting the social meaning and use of language in ontology concept formulation.

## 2　Acknowledging and Interpreting Social Discourse and Social Realism

In order for the social realism of the actors in a social setting to be captured, the perspectives of each actor needs to be acknowledged and incorporated into the reasoning behind the concept formulation process. As previously noted in Keen, Milton & Keen (2012a), developing an ontology to support a social process requires conceptualisation of the domain, and the influence of skills and perspectives of actors in the processes to be considered. In this research a complex social setting was selected to provide the context for consideration, development and operalization of a rigorous concept formulation methodology. Discourse from recordings of a community festival's voluntary management committee meetings were analysed and an ontology developed, which was grounded in the discourse of these meetings. The management meetings provided a rich source of text for concept formulation. The text of the meetings provided a way of understanding the social processes involved in the management and running of community events. The committee brought a broad range of skills and knowledge. There has been a relatively high turn-over of members of the committee over the past twelve months, which is a common concern in volunteer associations where there is a single focus on activities (Smith, 1994). This further highlights the need to continually share knowledge and language between the actors and offerings of the festival.

We have previously argued that actors will assume fundamentally different perspectives, based on their background and any formal or assumed roles in that setting (Keen, Milton & Keen, 2012b). Further, since perspectivism can provide a useful theoretical basis and is not incompatible with a common-sense realist stance taken in the coding steps (Keen, Milton & Keen, 2012b), the identification and clarification of perspectives has remained problematic to operationalize. Discourse analysis (Martin & Rose, 2008; Halliday, 1994) provides a structured framework to assist the researcher in moving from specific terms to establishing the meaning of sentences and multiple sentences. This is achieved through clarification of themes, rhymes, fields, tenor and genres – the stages of discourse analysis, as identified by (Martin & Rose, 2008).

We have applied a syntactic analysis method, with a philosophical lens of *common-sense realism* to make sense of the syntactic structure of sentences. We consider the common-sense world as modular in nature based on cultural driven target domains. In this paper we consider how the use of metaphor and stories are used in social settings to bridge the modular gaps and separations between actors' target domains.

The steps / stages involved in concept formulation were previously published in initial form in Keen, Milton & Keen (2012a) and revised in Keen, Milton & Keen (2012b). In 2013 we published the ontology concept formulation process (Keen, Milton & Keen, 2013b) which applied the rigorous interpretive process and techniques offered by discourse analysis. These papers describe an ontology concept formulation method that progressively moves from the specific term level (stages 1-3), to consider generalised relations (stage 4), to identification of perspectives evident in the text (stage 5), and then to establish the ontological structures that emerge from analysis of the setting, through the identification of patterns in the discourse, e.g. through actors' use of social metaphors and genres (stage 6).

## 3　The Validity of Ontology Concepts – Perspectives, Genres and Metaphors

Interpretive research poses as epistemological assumption that knowledge about the world is acquired through social constructions, such as language, consciousness, and shared meanings. The focus is on human sense making of the situations as they occur and on the meanings people assign to the situations. There are two strands of thought of interpretive research, based on this epistemological assumption – one is based on language and its meaning; the other is more related to phenomenology and hermeneutics (Klein & Myers, 2001).

In this paper we have extended the conceptual questions indentified in Cardoso & Ramos (2012) and refer to the interpretive methodology tools and logic required for rigorous and interpretive ontology concepts, in particular the issue of unpacking the complexity of social discourse and culture and how patterns in discourse, specifically the references to genres and metaphors, are used to bridge the interpretive gap between discourse.

Knowing what perspectives exist and how actors span perspectives gives us a way to more deeply understand what different ontologies may be required, and further, gives us an insight into how to bridge between reference ontologies.

Multiple perspectives are shown through patterns of fields, for example, two or more field may be simultaneously discussed and the interface between the fields negotiated as part of the discourse. However, other patterns of language use may also betray perspectives. Genres may help actors from other perspectives better understand the perspective of the speaker. Further, as identified by Pinker (2010), repeated vagueness or ambiguity in language was used, and often is useful in determining intentionality in social discourse.

An actor's use of modes, metaphors and genres, provide insight into how that actor attempts to express their perspective, while also attempting to appeal to the perspectives of others. It has been identified in this study that actors employ metaphors to cross the conceptual boundaries between domain perspectives. The use of metaphors and idioms provides a link between the referent concepts and intent or perspective of the speaker. An example is the use of a conventional metaphor *'I'm sure it costs an arm and a leg'*, which indicates that the actor is attempting to create a bridging reference between the perspectives of their 'experience' and a 'resource and planning' domain.

**The use of metaphors** is not any particular word or expression, it is the ontological mapping across conceptual domains, from the source domain to a target domain. The metaphor is not just a matter of language, but of thought and reason. The mapping is primary, in that it sanctions the use of source domain language and inference patterns for target domain concepts (Lakoff, 1992).

**The uses of genres** are often a narration or story, enacting social processes within a particular social context Martin & Rose (2008). Examples of genre are a timeline-based recounting of events as a narrative, providing a reference description of a given thing or event in time.

Genres as described by Martin & Rose (2007) are "staged, goal oriented social processes. Staged, because it usually takes actors more than one step to reach goals and goal orientation, because we feel frustrated, if we feel there is misunderstanding or if we don't accomplish the final steps in a goal process / strategy". Genres as generally adopted in discourse to reinforce a goal or social action. Actors refer to genres to reinforce and recount story or refer to an example, or provide a historical account, to reinforce policy with a divergent community or cross section of actors. An example of a genre is a timeline-based recounting of events as a narrative, as it provides a detailed description of a given thing or event in a time-less manner.

It is evident in discourse, unlike text, there are shifts in genres, as they are descriptive techniques to reinforce a configuration of meaning or the enactment of a social practice (e.g. in the form a reference story). Genres may be used by actors to provide a generalized account, historical account or a specific reference to an existing policy. The variations in a genre may be references to participants, tense / modality, the reference to activity sequences, attitudes, causes, phases or a recount of a time sequence.

In linguistic terms, genres are recurrent configurations of meaning and these configurations of meaning enact the social practices of a given culture. We consider that genres are significant in unpacking the critical points of a process through linguistic illustrations. Common understanding, as previously mentioned may not be achieved by an actor's reference to conduit metaphors, but instead by the intent and use of terms in expressions or non-conduit metaphors. It may be references to a limited number of common concepts, which creates the cohesion between an actor's expressions, within a specific genre of expression.

## 4   Interpreting the Layers and Contextualisation of Discourse

*All human understanding is achieved by iterating between considering the interdependent meaning of parts and the whole that they form. This principle of human understanding is fundamental to all the other principles (Klein and Myers, 1999).*

We consider the interpretation of layers and cotextualisation of discourse to be fundamental to ontology concept formulation when considering the mereology of a context, and fundamental in the common-sense realism approach to ontology concept formulation as adopted at all stages of the ontology concept formulation method (Stages 1- 6) published in Keen, Milton & Keen (2013b).

The objective of contextualization is to be faithful to the case and context. We do this by active observation and reflection of the functional use of language or discourse, with consideration of how actors use terms, sentences and references to fields, genres and metaphors.  In particular we believe the principle of contextualization is fundamental to the data collection and discourse analysis techniques adopted in stages 3 - 6 of the ontology concept formulation method.

We detail specifically Stages 5 and Stage 6 of the ontology concept formulation methodology below to illustrate the value of interpreting genres and metaphors in gaining a greater understanding of the use of linguistic terms in reference to the context and distinct actors, and perspective target domains present in the context.

**Stage 5: Interpreting the influence of context and culture: The pragmatic flow of discourse**

The objective of Stage 5, following on from the identification of the theme of the sentence at the term level (Stage 2), is to identify the field (what is being spoken about) at the multiple sentence level. The identification of fields in the discourse provides a mid-way categorization between the specifics of sentence-level meaning and the much broader idea of perspective. It does this by identifying the meaning and intent of multiple sentences. It has been identified by Martin & Rose (2008) that the flow in discourse is inherently influenced by an actor's existing knowledge, their social relationships or affiliations within the context, and the formal and social roles the actor(s) adopt. Therefore, this stage also includes the application of discourse analysis that gives rigor to discussing the complex interplay between the social context, culture, processes and social relationships evident in the text. This goes well beyond the term-based understanding of the context in (Stage 2), and does so in an integrated way. This is partly achieved by identifying tenors, fields and genres within the discourse, as these indicate formal and informal social relations between actors. This is then followed by the identification of the formal and informal roles, relationships and affiliations of the speakers.

The objective of Stage 5 has been to interpret the context and culture surrounding the use and interpretation of language and the social relationships influence the use of language, for instance the use of 'jargon'. Multiple actors contribute to the discussion, interpreting the influence of context and culture and the pragmatics of the discourse, as they consider the implications of the various fields.

**Stage 6: Identifying perspectives through patterns in discourse**

This stage assists in recognising perspectives by seeking patterns in discourse. Specifically, this stage aims to clarify how actors use language and language tools in a way that indicates perspectives and span perspectives. We have identified that discourse is heavily dependent on metaphors, which are non-literal, but meaningful within the context of a discussion. Patterns we have found include:

(1) The recurrence of fields and the relationship of fields with specific actors or roles
(2) How the genre of communication is used in the discourse to relate to other actors. Both of these allow us to see perspectives in the discourse.
(3) How metaphors and the use of genres are used to bridge the gap between fields, actors' roles and the spanning of perspectives.

Stage 5 and 6 of the ontology concept formulation methodology address the need to identify multiple interpretations. These stages provide a deeper understanding of how actors use discourse tools, such as genres and metaphors, provide an indication of an actor's use of language.

## 5    Illustration of the use of Metaphors and Genres

*We recognise the need to be sensitive to possible differences in interpretations among the participants as are typically expressed in multiple narratives or stories of the same sequence of events under study. (Klein and Myers, 1999)*

The following section will provide an example of an illustrative fragment of text, which has been gathered from a community event management meeting discourse. This fragment of discourse was gathered during the operational phase of the event management life cycle.

We recognise that multiple concurrent interpretations exist in social complex settings. We recognise that natural language has a semantic structure and that fields, genres and metaphors provide insight into the semantic and merelogical structure of language. We believe the process of recognition of multiple interpretations in concurrent discourse is the process of sense making or semantic interpretation of discourse.

This fragment has been chosen because of the evidence of multiple actors, multiple roles and perspectives present. We interpret the, the influence of the actors' role, the actors' perspectives and most importantly for the purpose of this paper, actors' use of genre and references to metaphors. Fragments are analysed multiple times to unpack the true interpretation of each term, sub-sentence and sentence in context, considering the intent and context in which the language is used.

**Fragment 1: 'Keep the local artists on side'**

This fragment is discussing the broad field of a motion of a meeting, in reference to the delivery and judgement of the '*design and production of marketing material*'. More specifically this fragment is discussing:

- Artwork being judged / reviewed as part of a local school completion
- The value of supporting the local community and artists and being supported by the community in the delivery of this program.
- There is a query about the process of review and the process of judgment.
- There is talk of timelines and urgency.
- There is some tension

This illustrative fragment was gathered at the 'Operational Planning' phase of the event management life cycle. This fragment is discussing the process of conducting a local school art competition, to include local children in the branding of the festival. The competition would be judged and the winner's artwork would be selected as the branding image of that year's festival.

As we review this fragment we will consider the elements of the fragment at the sentence level, considering the actors' present, the use of language, and most importantly the use of metaphors and genres and how they are used to communicate intent and bridge perspectives.

> **1.1 School Liaison and Local Performer:** Before we ask anyone we should keep the local artists on side.
> **1.2 Secretary:** I think we do, definitely.
> **1.3 President:** Well within two weeks we will have all the entries together, and who wants to judge them?
> **1.4 Artistic Director:** That's a good question, we should possibly have … there is a priority in getting it judged and decided on. Last year, I'm sure we were still dilly-dallying around in September weren't we?

The following table (Table 1) provides a simple analysis of the fragment at the sentence and sub-sentence levels to consider the contribution of the actors' roles, perspectives and the use of metaphors.

| Fragment sentences | Actor's Role | Perspectives | Metaphor | Reference to Genres |
|---|---|---|---|---|
| 1.1 Before we ask anyone we should keep the local artists on side. | School Liaison and Local Performer | Planning / Resource, Service, Process and System, Experience | "keep the local artists on side" | |
| 1.2 I think we do, definitely. | Secretary | | | |
| 1.3 Well within two weeks we will have all the entries together, and who wants to judge them? | President | Planning / Resource, Regulatory / Governance, Resource / System | | |
| 1.4 That's a good question, we should possibly have … there is a priority in getting it judged and decided on. Last year I'm sure we were still dilly-dallying around in September weren't we? | Artistic Director | Resource / Planning, Process, Experience. | "dilly-dallying" | "Last year I'm sure we were still dilly-dallying around in September weren't we?" |

**Table 1:** Evidence of Metaphors and Genres

We believe that to cross perspectives, metaphors are essential for creating blends between imagery, with the intent to create common objective and goals within a complex social setting. We consider the above metaphors and their literal meanings vs. contextual meanings, and how metaphors, when referred to in context, can reinforce intent and form imagery.

**Line 1.1 Metaphor - *keep the local artists on side***

*Literal meaning of metaphor line 1.1 metaphor – 'on-side' - Helping or giving an advantage.*

The context in which the statement was made has greater weighting and imagery then the literal meaning of this statement, as this metaphor refers to the topic of local and community. This topic is of greater significance in the context of this discussion, as the primary objective of this context is the delivery of a community event which has a strong common value of community inclusiveness and artistic direction. The concepts of 'local' and 'artists' add greater value to the metaphoric term 'on-side'.

It is believed that the conceptual inclusion of the terms 'local' and 'artists' also creates valuable community imagery for actors.

**Line 1.4 Metaphor – * Last year I'm sure we were still dilly-dallying around in September weren't we?***

> *Literal meaning of metaphor line 1.4 'dilly-dallying' - Waste time through aimless wandering or indecision.*

In reference to this context, this is an important metaphor, as it indicates the actor's reference point of time and the value of time, while also indicating the need for a process, common activity and common decision to be made by multiple actors.

Line 1.4 refers to a metaphor and genre. It is believed that the genre is included by the actor to create a reference to a time point, time period, previous experience and the value of time.

The use of a genre in this context has provided an end-time point in the overall event management lifecycle. The use of the genre has also shaped expectations, constraints and influences the reference to previous knowledge, while also creating communal knowledge and focus.

## 5.1 Metaphors, Genres and the Bridging target domains in complex communicative interactions

*Do we systematically use inference patterns from one conceptual domain to reason about another domain? (Lakoff & Johnson, 2003)*

We have identified patterns and correspondence between unique domains. We have identified that metaphors and genres are linguistic tools to metaphorically map between domains, and that two domains correlated if concept mapping is established by actors. We recognise in the broader domain of event management, that multiple target domains correlate and the correlation is not purely abstract, but determined by the actors' reference domains and inherent perspectives.

**Line 1.4 Genre - *Last year I'm sure we were still dilly-dallying around in September weren't we?***

Line 1.4 has provided a metaphor and reference to a genre as the actor aims to enforce and influence negotiation between a social setting, with the influence of actor experience, and an inherent reference to time points, system or time-lines in the committee's activities and processes.

If we consider the patterns in the use of metaphor and genres and the contextual factors, which influence the use of metaphors and genres, then we can gain greater insight into the use of metaphorical language in complex linguistic contexts. We are most interested in how metaphors and genres allow actors to bridge source and target domains and how key concepts create weighting, and communal knowledge, where actors' perspectives may influence common or distinct imagery.

We believe that the use metaphors influence social processes, communal knowledge and, most importantly, create common goals in social processes. We believe there is significant value in a grounded conceptual analysis methodology, as it provides greater insight into the patterns, use and conceptual purpose of discourse when analysed in context.

# 6   Conclusion

*Nietzsche suggests that metaphors and models need to be employed to articulate the perspectives adopted by actors, and that these perspectives are largely framed by the use of these metaphors and models (Honderich, 1995:622).*

We have illustrated a methodology that bridges the gap between term identification and class membership conditions, based on common-sense realism. This paper recognises that perspectivism influences the flow of discourse, and proposes that perspectivism, and the use of metaphors and other patterns provide the linguistic basis for achieving ontological modularity. The process of division of an ontology into modules (ontological modularity) relies on the selection and definition of modules that are self-consistent, share a common goal or goals, and express the purposes inherent in the specific perspective (Parent & Spaccapietra, 2008). By identifying the commonalities between the perspective based modules we can attempt to bridge the gap between multidisciplinary cultures to identify the commonalities, so that one's world or frame of reference may be broadened.

It has been identified that the use of perspectivism clarifies the nature of intention and interchanges in social discourse and the roles and goals being adopted by the participants, and hence the nature and the boundary of the discourse, or the interplay between actors. This also informs the nature of the boundaries between the divergent ontologies.

The uses of metaphors provide the intentional labels for objects, thoughts, actions, ideas and concepts. Actors' use of metaphors provides insight into the perspectives adopted by the actors, and how actors' perspectives facilitate communication with fellow actors. We consider the value of perspectives in providing linguistic mechanisms for modularising ontologies, which are aligned with the philosophy of common-sense realism.

Specifically, we consider actors' use of language as a tool to span perspectives and how discourse analysis tools and techniques enable a deeper interpretive understanding of the layers of discourse, when derived from a rich context. The purpose of this study is to illustrate and richly interpret the patterns we have found in the use of discourse analysis, from linguistic, natural language and pragmatic perspectives. This detailed method is the significant contribution of this study, as it provides the logical steps to interpret the meaning of language and actors' abilities to reference the use the metaphors and genres to bridge the gap between their distinct perspectives.

## Copyright